\documentclass{PoS}
\usepackage{graphicx}
\newcommand{\bo}{\raise-1mm\hbox{\Large$\Box$}}              % D'Alembertian
\PoS{PoS(LAT2005)149}
\title{Two Color QCD beyond the BEC r\'egime}
\ShortTitle{Two Color QCD}
\author{\speaker{Simon Hands}\thanks{PPARC Senior Research Fellow}\\
University of Wales Swansea\\
E-mail: \email{s.hands@swan.ac.uk}}
\author{Seyong Kim\\Sejong University, Seoul\\
E-mail: \email{skim@sejong.ac.kr}}
\author{Jon-Ivar Skullerud\\Trinity College, Dublin\\
E-mail: \email{jonivar@maths.tcd.ie}}
\abstract{We present results of simulations of Two Color QCD using two flavors
of Wilson quark in the fundamental representation, at non-zero quark chemical
potential $\mu$, on an $8^3\times16$ lattice. Results for the quark number
density, quark and gluon energy densities, and superfluid condensate
are qualitatively distinct from the behaviour expected on the assumption that
the dominant degrees of freedom are tightly bound scalar diquarks which Bose 
condense;
rather the scaling with $\mu$ is more suggestive of a 
Fermi surface disrupted by a Cooper pair condensate. We also present evidence
both for screening of the static potential, and color deconfinement, arising
solely as a result of a non-zero quark density.}

\FullConference{XXIIIrd International Symposium on Lattice Field Theory\\

                 25-30 July 2005\\

                 Trinity College, Dublin, Ireland}
\begin{document}
\section{Preamble}
QCD with gauge group SU(2) and quark chemical potential $\mu\not=0$
has enjoyed much recent interest \cite{Simon}. It lies in a 
class of models with long-ranged interactions (together with QCD with 
non-zero isospin chemical potential (ie. $\mu_u=-\mu_d$) and vectorlike 
theories with adjoint
quarks) in which the quark determinant
$\mbox{det}M(\mu)$ is real, and hence amenable to study using standard LGT
algorithms. 

In Two Color QCD, 
$q$ and $\bar q$ fields live in equivalent representations of the 
color group SU(2), implying that chiral multiplets contain both $q\bar q$ mesons
and $qq$ baryons. In the chiral limit where $m_\pi\ll m_\rho$, where $\rho$
generically denotes any non-Goldstone hadron, it is possible to study the 
$\mu$-dependence of the model systematically using chiral perturbation theory
($\chi$PT)
\cite{KTSVZ}. The key result is that the ground state has non-zero quark
density $n_q>0$ for all $\mu$ greater than some onset value 
$\mu_o={1\over2}m_\pi$. At the same point a superfluid order parameter
$\langle qq\rangle\not=0$ develops, signalling spontaneous breakdown of the 
original global U(1) baryon number symmetry. Since the transition is second
order the matter, consisting of tightly-bound $qq$ scalars, 
may be arbitrarily dilute, and in the limit
$\mu\to\mu_{o+}$ is a textbook Bose-Einstein Condensate (BEC). In this 
r\'egime the quantitative predictions of $\chi$PT read: 
\begin{equation}
n_q\propto f_\pi^2(\mu-\mu_o);\;\;\;\langle
qq\rangle\propto
\sqrt{1-\left({\mu_0\over\mu}\right)^4}
\;\;\;\;\;\Rightarrow\lim_{\mu\to\infty}\langle qq(\mu)\rangle=\mbox{const.}
\label{eq:chiPT}
\end{equation}
This behaviour has indeed been confirmed by simulations with staggered fermions
\cite{AADGG,adj1,KSHM}

This picture should be contrasted with another paradigm for superfluidity, 
namely BCS condensation of weakly interacting quark Cooper pairs from opposite
points of a Fermi surface with radius $k_F\approx E_F=\mu$. For such a system,
we obtain $n_q$ simply by counting states within the Fermi sphere, and the
condensate by assuming that only states within a layer of thickness
$\Delta\ll\mu$
around the Fermi surface participate in the pairing:
\begin{equation}
n_q\propto \mu^3;\;\;\;\langle qq\rangle\propto\Delta\mu^2.
\label{eq:BCS}
\end{equation}
The same rule of thumb predicts the quark energy density
$\varepsilon_q\propto\mu^4$.

We might expect the crossover from BEC to BCS for 
$(\mu-\mu_o)\gg{1\over2}(m_\rho-m_\pi)$, 
at which point the Goldstones are no longer
distinguished hadrons, and the fermionic nature of the constituents comes into
play. The chiral limit is thus unimportant, and for
this reason we have chosen to revisit Two  Color QCD using Wilson fermions 
\cite{wil1}. Pioneering simulations by the Hiroshima group appeared in
\cite{Hiroshima}. As shown below, an advantage of the Wilson formulation
is that simulations with $N_f=2$ quark flavors are possible,
ensuring the theory is asymptotically free and confining for $\mu=T=0$
for all couplings, with a controllable continuum limit as $\beta\to\infty$.
This makes the Wilson LGT particularly suitable for a study of gluodynamics in
the presence of a background baryon charge density, though it should be stressed
that in contrast to 3-color QCD, 
there is here no physical distinction between fundamental
and anti-fundamental charges.

\section{Formulation and Simulation}

The $N_f=2$ quark action we want to simulate is as follows:
\begin{equation}
S=\bar\psi_1M\psi_1+\bar\psi_2M\psi_2
-J\bar\psi_1(C\gamma_5)\tau_2\bar\psi_2^{tr}
+\bar J\psi_2^{tr}(C\gamma_5)\tau_2\psi_1,\label{eq:action}
\end{equation}
where flavor indices are written explicitly, and we have included scalar
isoscalar diquark source terms, which will help to reduce IR fluctuations in the
superfluid phase \cite{KSHM}. The action coincides
with that of \cite{wil1} with the identification
$J=2\bar\jmath$, $\bar J=2j$.
The charge conjugation matrix $C$ defined by
$C\gamma_\mu C^{-1}=\gamma_\mu^*$ with $\gamma_\mu^*=\gamma_\mu^{tr}$ in
Euclidean metric has the properties $C^\dagger=C^{-1}=-C$, $[C,\gamma_5]=0$.
The crucial Two Color identity is $\tau_2U_\mu(x)\tau_2=U_\mu^*(x)$.
The source strengths $J$ and $\bar J$ are
{\it a priori} independent and their signs are arbitrary for now. The matrix
$M$ is the textbook Wilson fermion operator
\begin{equation}
M_{xy}(\mu)=\delta_{xy}
-\kappa\sum_{\nu}[(1-\gamma_\nu)e^{\mu\delta_{\nu0}}U_\nu(x)\delta_{y,x+\hat\nu}
+(1+\gamma_\nu)e^{-\mu\delta_{\nu0}}U_\nu^\dagger(y)\delta_{y,x-\hat\nu}]
\end{equation}
with properties 
\begin{eqnarray}
\gamma_5M(\mu)\gamma_5&=&M^\dagger(-\mu)\label{eq:dagger}\\
C\tau_2M(\mu)C^{-1}\tau_2&=&M^{tr}(-\mu)\label{eq:transpose}\\
\Rightarrow\;\;\;\;
(C\gamma_5)\tau_2M(\mu)(C\gamma_5)^{-1}\tau_2&=&M^*(\mu)\label{eq:star}
\end{eqnarray}
Property (\ref{eq:star}) implies $\mbox{det}M(\mu)$ is real, but because
$M$ contains both hermitian and antihermitian non-constant
components there is no
proof that it is positive. This confirms the impossibility of simulating a
single Wilson flavor without a Sign Problem, and also suggests that even
for $N_f$ even there is
an ergodicity problem along the lines of that found for adjoint staggered
fermions \cite{adj1}, where changing the sign of $\mbox{det}M$ requires an
eigenvalue to flow through the origin, which the Two-Step Multi-Bosonic 
algorithm 
can manage but the Hybrid Monte Carlo (HMC) algorithm not.
                                                                                
Now, with the change of variables $\bar\phi=-\psi_2^{tr}C\tau_2$,
$\phi=C^{-1}\tau_2\bar\psi_2^{tr}$ and the dropping of the index from
flavor 1, eq. (\ref{eq:action}) can be recast as:
\begin{equation}
S=(\bar\psi,\bar\phi)\left(\matrix{ M(\mu)&J\gamma_5\cr
                           -\bar J\gamma_5&M(-\mu)\cr}\right)
\left(\matrix{\psi\cr\phi\cr}\right)\equiv\bar\Psi{\cal M}\Psi.
\label{eq:bilinear}
\end{equation}
Note that the action is bilinear in the variables $\Psi,\bar\Psi$, so that the
Grassmann integral yields a factor $\mbox{det}{\cal M}$ rather than a Pfaffian.
Using the identity
\begin{equation}
{\rm det}\left(\matrix{X&Y\cr W&Z\cr}\right)=
{\rm det}X{\rm det}(Z-WX^{-1}Y)
\end{equation}
and properties (\ref{eq:dagger},\ref{eq:transpose}) we deduce
\begin{equation}
\mbox{det}{\cal M}=\mbox{det}(M^\dagger(\mu)M(\mu)+J\bar J).
\end{equation}
Hence positivity of $\mbox{det}{\cal M}$ requires the product $J\bar J$ to be
real and positive, which translates into the requirement that the diquark
source term be antihermitian \cite{wil1}.
Since no eigenvalue of
${\cal M}$ can vanish, the ergodicity problem is also cured.
                                                                                
Now use
(\ref{eq:bilinear}) to write
\begin{equation}
{\cal M}^\dagger{\cal M}=
\left(\matrix{M^\dagger(\mu)M(\mu)+\vert\bar J\vert^2&\cr
&M^\dagger(-\mu)M(-\mu)+\vert J\vert^2\cr}
\right).
\end{equation}
The off-diagonal terms can be shown to vanish if $\bar J=J^*$
using (\ref{eq:dagger}); moreover
the same identity applied to the lower block yields
\begin{equation}
\mbox{det}{\cal M}^\dagger{\cal M}=[\mbox{det}(M^\dagger(\mu)M(\mu)+\bar
JJ)]^2.\end{equation}
It is therefore possible to take the square root analytically, 
by using pseudofermion fields 
with weight $(M^\dagger M+\vert J\vert^2)^{-1}$.
This has the advantage of {\em(a)}
requiring
matrix/vector multiplications of the half the dimensionality, and {\em(b)}
permitting
a hamiltonian evaluation and hence the use of an exact HMC algorithm.
It is equivalent to the even/odd partitioning step used for staggered fermion
gauge theories, but is more transparent since all lattice
sites are physically equivalent, making the force term easier to implement.
The trick was used in \cite{KSHM}, though because the staggered version
still requires a Pfaffian rather than a determinant, an HMD
algorithm was used in that case.

\section{Observables}
                                                                                
The most fundamental is the quark density,
obtained as a derivative of the free energy:
\begin{eqnarray}
n_q\equiv-{1\over V}{{\partial\ln{\cal Z}}\over{\partial\mu}}
&=&{1\over{V{\cal Z}}}\int D\Psi D\bar\Psi
(\bar\psi,\bar\phi)\left(\matrix{{{\partial M}\over{\partial\mu}}&\cr
                         &-{{\partial M}\over{\partial\mu}}\cr}\right)
\left(\matrix{\psi\cr\phi\cr}\right)e^{-S}\nonumber\\
&=&\kappa\langle\bar\psi_x(\gamma_0-1)e^\mu U_0(x)\psi_{x+\hat 0}
+\bar\psi_x(\gamma_0+1)e^{-\mu}U_0^\dagger(x-\hat0)\psi_{x-\hat
0}\rangle\nonumber\\
&\phantom{=}&\!\!\!
-\kappa\langle\bar\phi_x(\gamma_0-1)e^\mu U_0(x)\phi_{x+\hat 0}
   +\bar\phi_x(\gamma_0+1)e^{-\mu}U_0^\dagger(x-\hat 0)\phi_{x-\hat 0}\rangle.
\end{eqnarray}
The apparently unphysical term 
$-2\kappa\langle\bar\psi(\partial_0-\mu)\psi\rangle$ is irrelevant in the long
wavelength limit. We have checked that at 
saturation $\lim_{\mu\to\infty}n_q=2N_fN_c$.
The quark energy density $\varepsilon_q$ is approximated by the 
same expression with a relative minus sign between forward and backward 
timelike links.
                                                                                
For the diquark condensate, it is convenient first to introduce
rescaled source strengths $\{j,\bar\jmath\}=\kappa^{-1}\{J,\bar J\}$, and then
the orthogonal combinations $j_\pm=j\pm\bar\jmath$. We then have
\begin{equation}
\langle qq_\pm\rangle\equiv-{1\over V}
{{\partial\ln{\cal Z}}\over{\partial j_\pm}}
={\kappa\over2}\langle\bar\psi\gamma_5\phi\mp\bar\phi\gamma_5\psi\rangle.
\end{equation}
Since the diquark condensate is not a component of a conserved current, its
normalisation is to some extent arbitrary: we prefer to normalise it with the
same factor of $\kappa$ as the quark density.

\section{Results}
We have studied an $8^3\times16$ lattice with parameters $\beta=1.7$,
$\kappa=0.1780$, using a standard Wilson plaquette action. Unfortunately our
results are not directly comparable with those of \cite{Hiroshima} due to 
their use of an improved gauge action. Studies of the $\mu=0$ string tension
yield a lattice spacing $a=0.220$fm, and the spectrum 
$m_\pi a=0.800(2)$, $m_\pi/m_\rho=0.920(3)$ \cite{wil1}. In simulations with
$\mu\not=0$ we have used a diquark source $ja=0.04$. So far we have accumulated
roughly 300 HMC trajectories of mean length 0.5 for $\mu a\in[0.3,0.9]$.

\begin{figure}[ht]
\begin{center}
\includegraphics[width=2.8in]{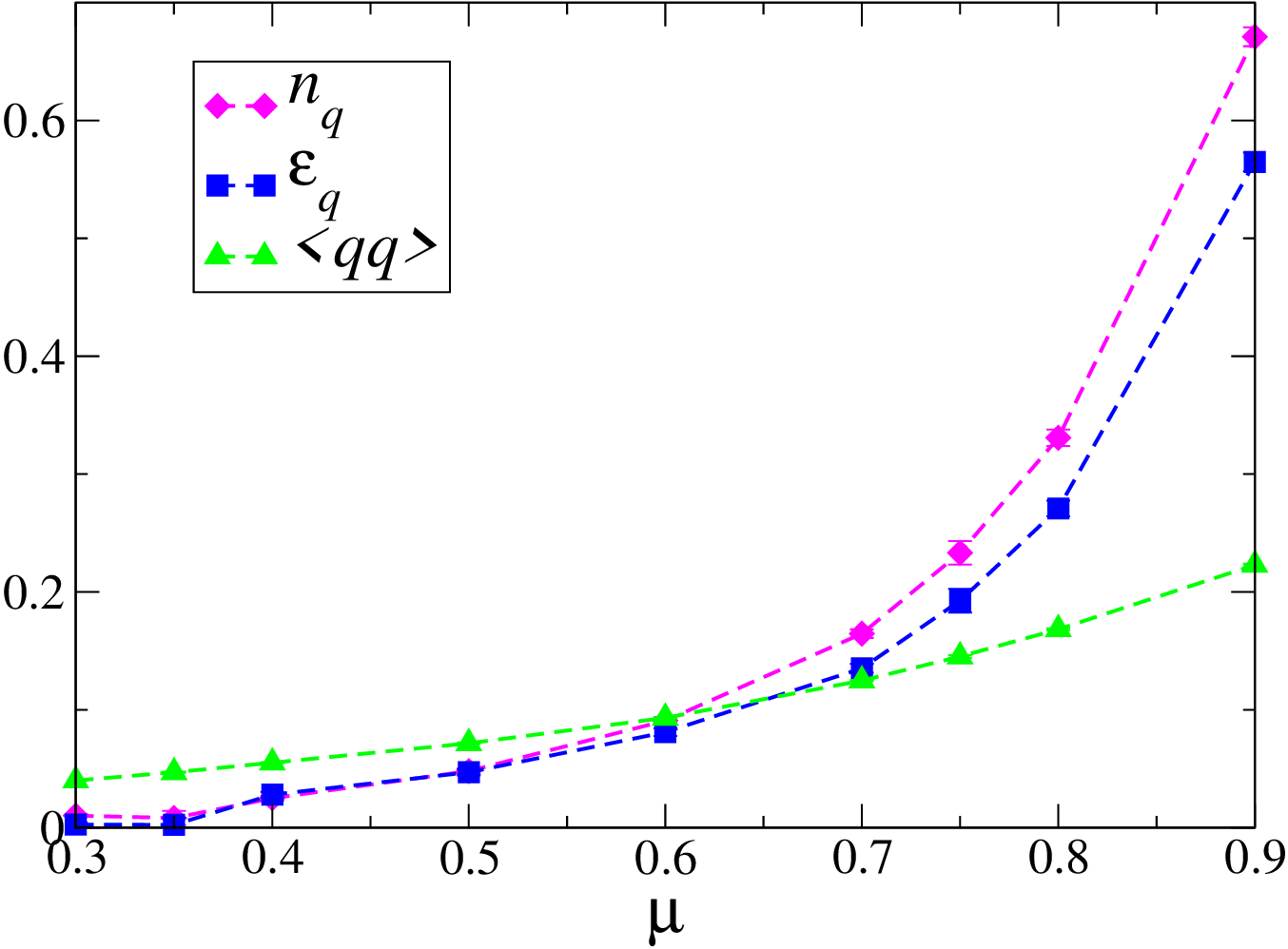}
\includegraphics[width=2.8in]{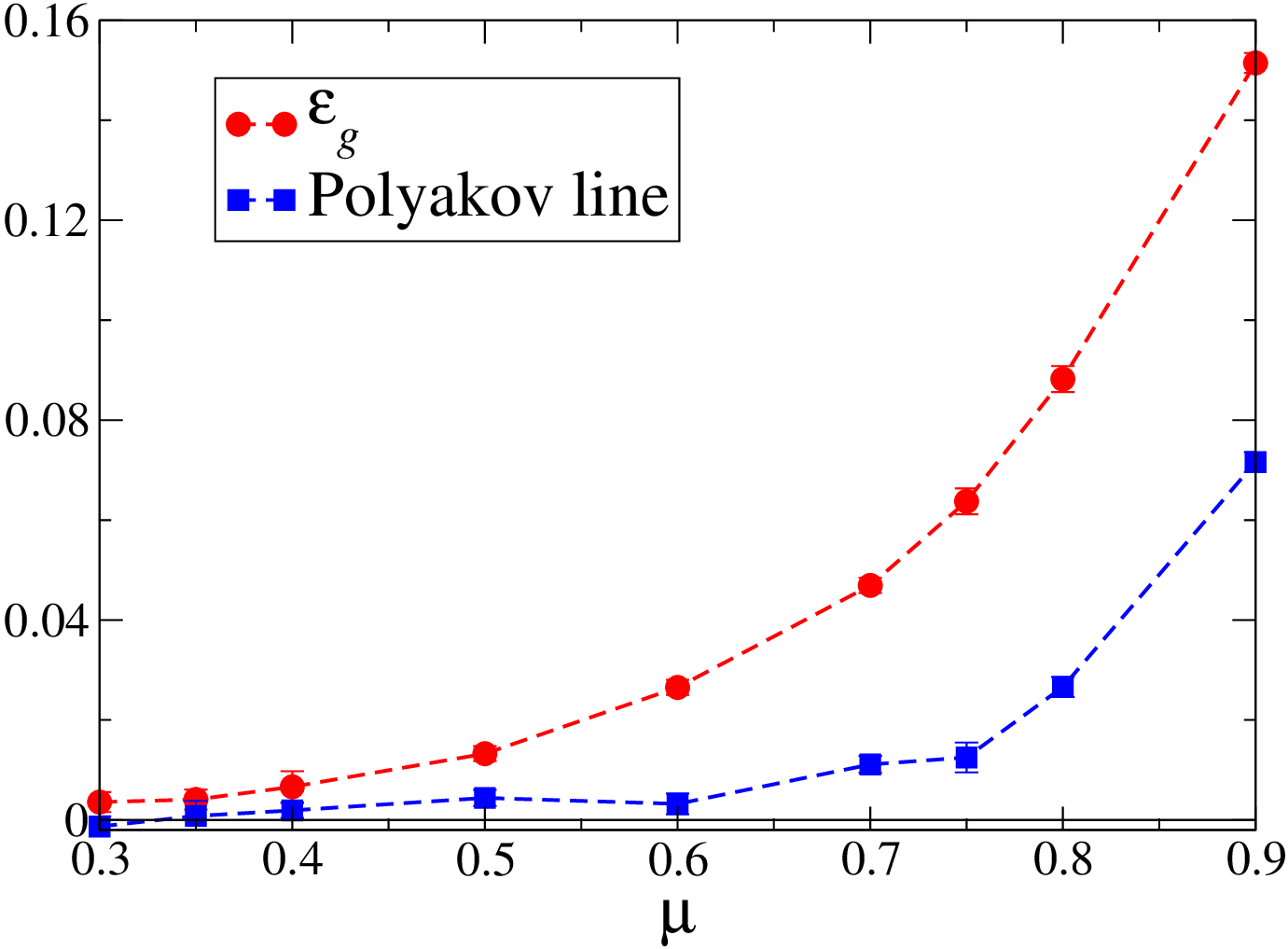}
\vskip -0.3cm
\caption{Quark observables\hskip 3.6cm {\bf Figure 2:} Gluon observables}
\end{center}
\vskip -0.3cm
\setcounter{figure}{2}
\end{figure}
In Fig.~1 we plot quark observables (with a suitable subtraction so that
$\varepsilon_q(\mu)=0$), and in Fig.~2 the gluon energy density
$\varepsilon_g={3\beta\over2}\langle\mbox{tr}(\bo_t-\bo_s)\rangle$ 
%(ie. ignoring
%Karsch corrections \cite{Karsch}) 
and Polyakov line, by far the
noisiest observable. 
The behaviour of the superfluid condensate $\langle qq\rangle$
vs. $\mu$ is 
qualitatively very different from the negative curvature seen in the 
$j\to0$ limit in studies with staggered fermions\cite{AADGG,adj1,KSHM}. 
Moreover the
magnitude  of $n_q(\mu)$ is much smaller, suggesting following (\ref{eq:chiPT})
that the effective $f_\pi$ is much smaller, and hence the hadron degrees of
freedom more strongly interacting.

\begin{figure}[ht]
\begin{center}
\includegraphics[width=2.8in]{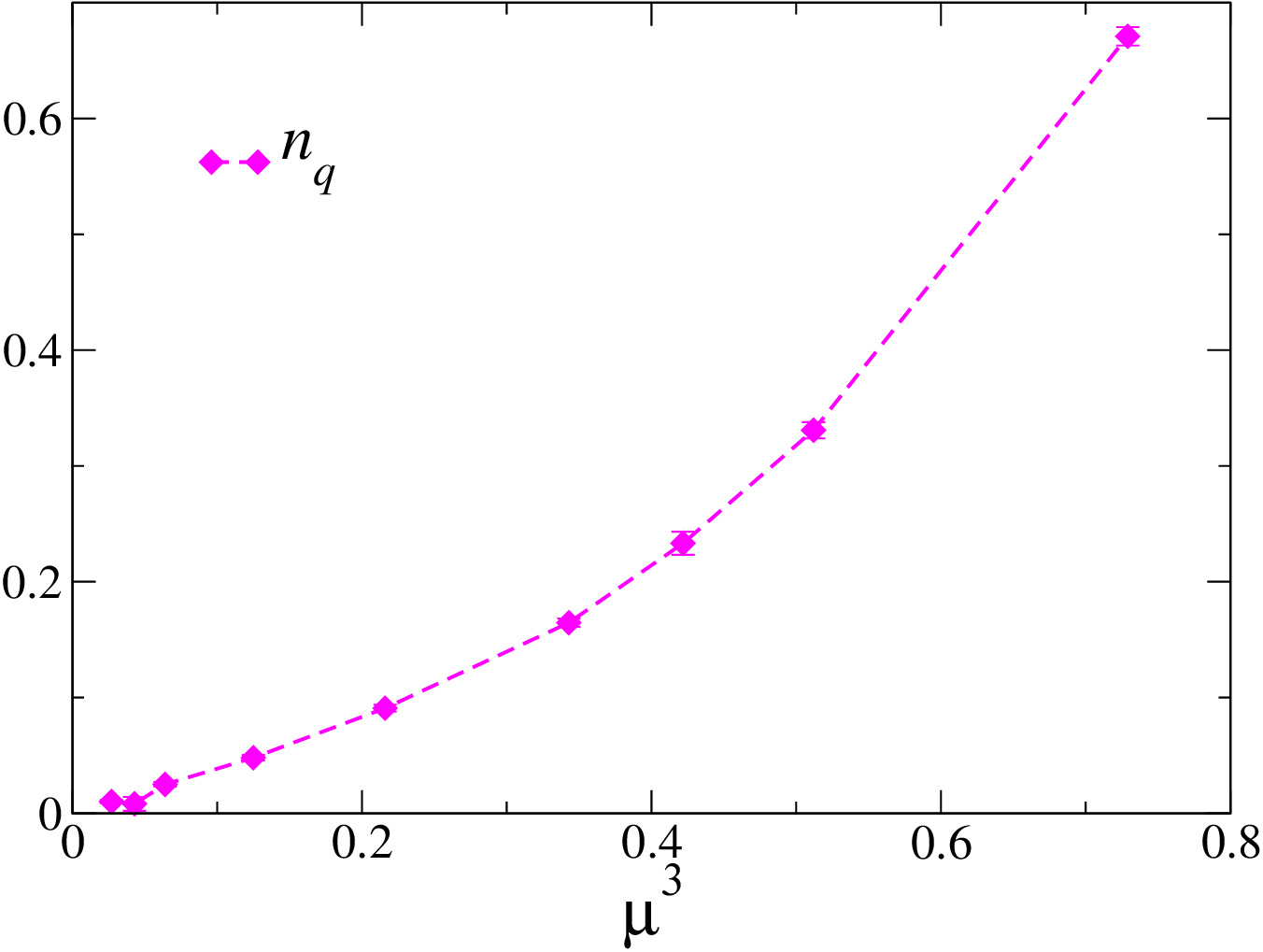}
\includegraphics[width=2.8in]{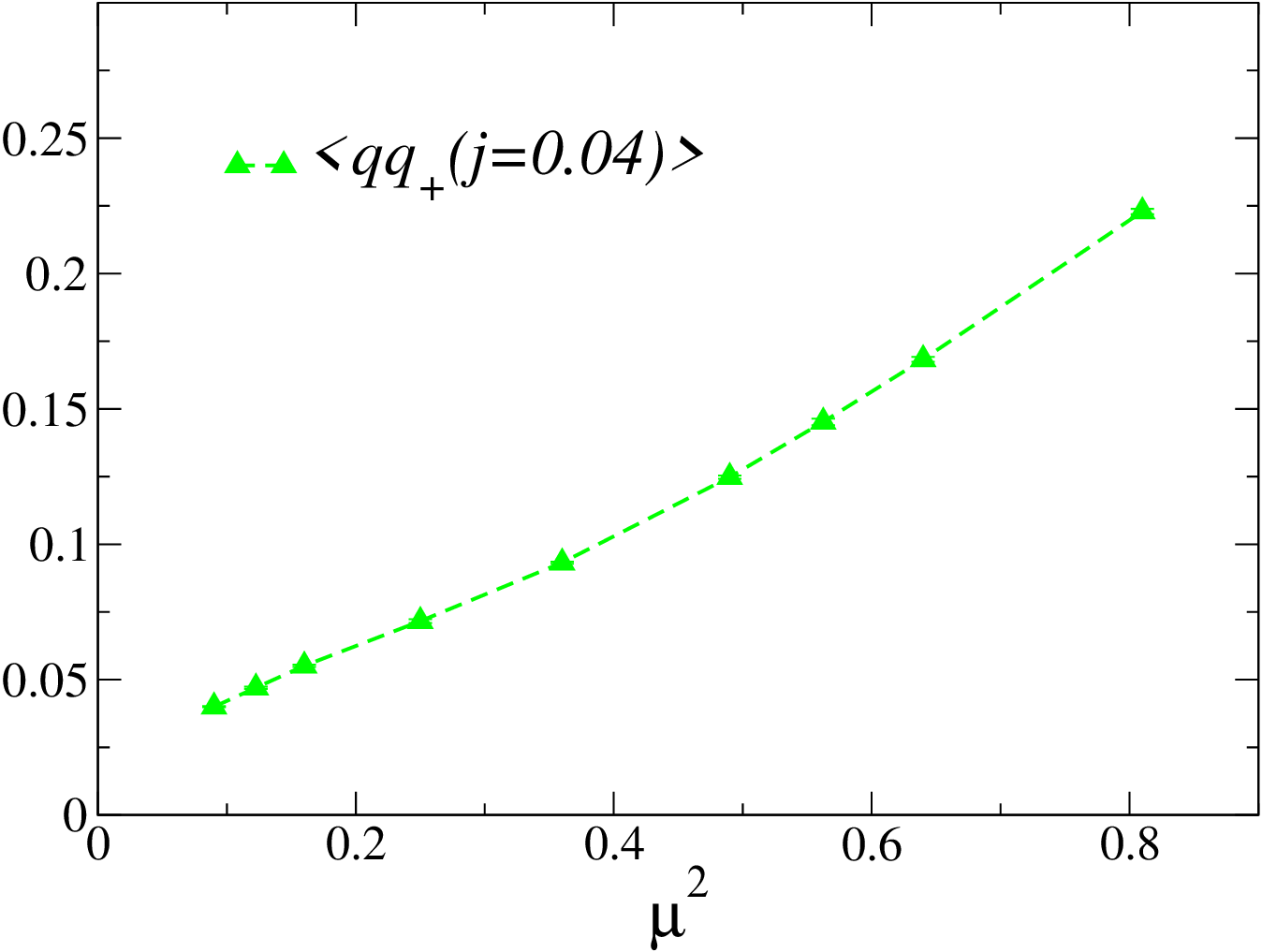}
\vskip -0.3cm
\caption{Quark number density vs. $\mu^3$\hskip 2.1cm {\bf Figure 4:} 
Diquark condensate vs.
$\mu^2$}
\end{center}
\vskip -0.3cm
\setcounter{figure}{4}
\end{figure}
This motivates us to replot the data following (\ref{eq:BCS}), as shown in
Figs.~3 \& 4. There is some evidence to support the formation of a Fermi sphere;
$n_q$ increases if anything more rapidly than $\mu^3$, although the impact of
lattice artifacts has still to be determined. Similarly, the
condensate is consistent with BCS behaviour being recovered in the
$j\to0$ limit, though data at smaller $j$ will be required to confirm this.
The most spectacular evidence 
for a Fermi sphere comes from the energy densities, 
plotted in Fig.~5. A significant fraction of the energy density, $O$(30\%),
comes from gluons, suggesting that the medium which forms is
strongly-interacting.
\begin{figure}[ht]
\begin{center}
\includegraphics[width=2.8in]{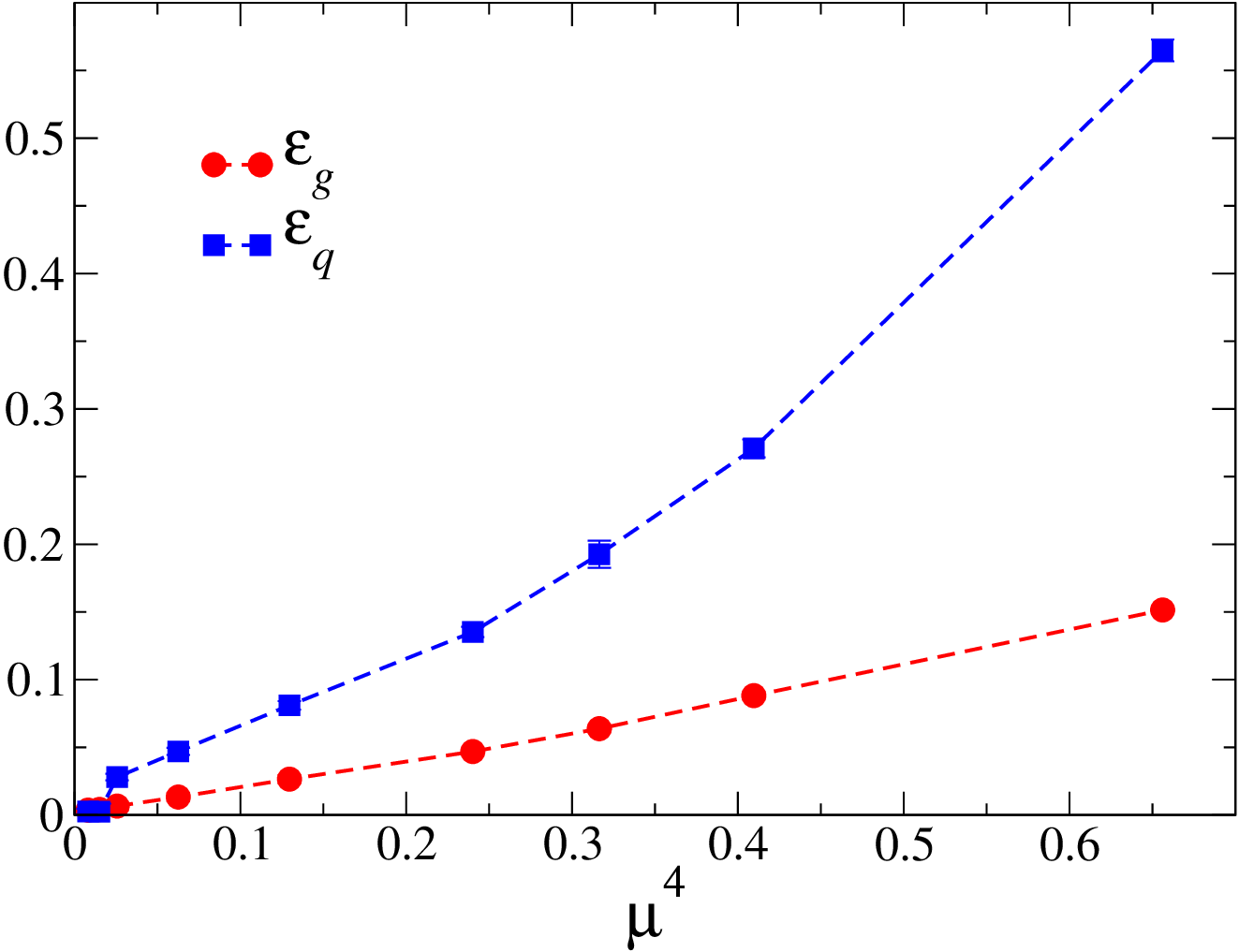}
\includegraphics[width=2.8in]{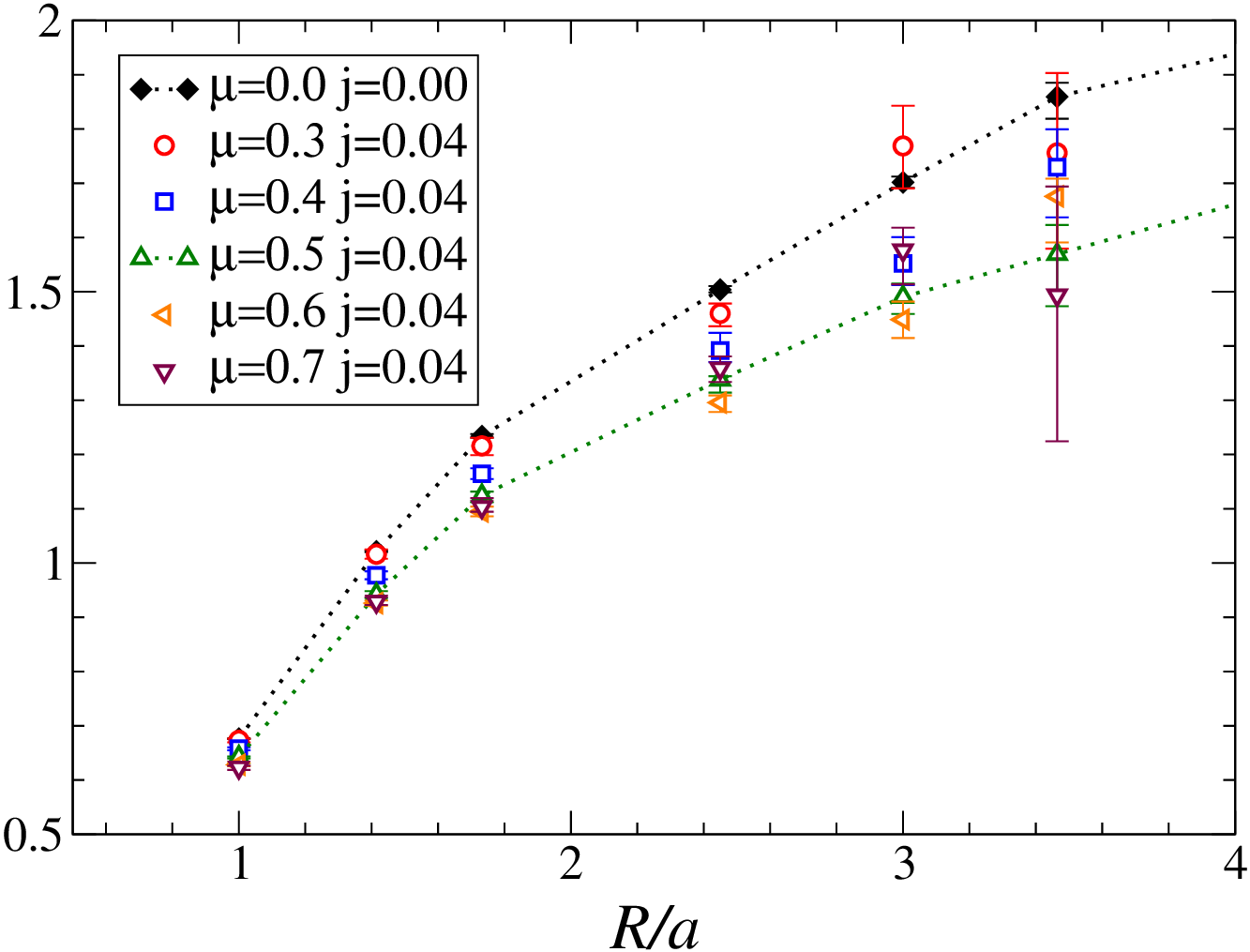}
\vskip -0.3cm
\caption{Quark and gluon energy densities vs. $\mu^4$\hskip 0.8cm
{\bf Figure 6:} Static potential $V(r)$ for various $\mu$}
\end{center}
\vskip -0.3cm
\setcounter{figure}{6}
\end{figure}
Finally in Fig.~6 we show the static quark potential $V(r)$,
showing clear evidence for screening for $\mu a\geq0.4$.

\section{Outlook}

The parameter set chosen for our initial study appear to place the Two
Color quark medium in a qualitatively different r\'egime from that described
by $\chi$PT and studied numerically using staggered fermions. If our guess that
the system has a Fermi surface disrupted by a diquark condensate is borne out by
further analysis, then the 
system may have a lot more in common with high density QCD than originally 
thought. One exciting possibility is that the onset transition predicted by
$\chi$PT at
$\mu_o a\simeq0.4$ does not coincide with the deconfining transition, which the
Polyakov loop data of Fig.~2 suggest may not occur until $\mu a \approx0.6(1)$.
Future study will also focus on the hadron spectrum, inspired by the evidence
for
decrease in mass for vector states reported in \cite{Hiroshima},
and a study of the gluon propagator.

\section*{Acknowledgements}
SK thanks PPARC for support during his visit to Swansea in 2004/05. 
We have also benefitted
from discussions with Kim Splittorff, and wish to warmly thank Shinji Ejiri and
Luigi Scorzato for their participation in the early stages of this project.

\end{document}